\documentclass[prl,aps,epsfig,twocolumn,showpacs]{revtex4}
\usepackage{epsfig,amssymb}

\newcommand\beq{\begin{equation}}
\newcommand\eeq{\end{equation}}
\newcommand\bea{\begin{eqnarray}}
\newcommand\eea{\end{eqnarray}}
\newcommand\non{\nonumber}
\newcommand\noi{\noindent}
\newcommand\al{\alpha}
\newcommand\be{\beta}
\newcommand\de{\delta}
\newcommand\ep{\epsilon}
\newcommand\ka{\kappa}

\newcommand\pa{\partial}
\newcommand\bib{\bibitem}

\begin{document}

\textheight=23.8cm

\title{\Large Gapless points of dimerized quantum spin chains: analytical and 
numerical studies}
\author{\bf V. Ravi Chandra, Diptiman Sen and Naveen Surendran} 
\affiliation{\it Centre for High Energy Physics, Indian Institute of Science,
Bangalore 560012, India}

\date{\today}

\begin{abstract}
We study the locations of the gapless points which occur for quantum spin 
chains of finite length (with a twisted boundary condition) at particular 
values of the nearest neighbor dimerization, as a function of the spin $S$ 
and the number of sites. For strong dimerization and large values of $S$, a 
tunneling calculation reproduces the same results as those obtained from more 
involved field theoretic methods using the non-linear $\sigma$-model approach.
A different analytical calculation of the matrix element between the two N\'eel
states gives a set of gapless points; for strong dimerization, these differ 
significantly from the tunneling values. Finally, the exact diagonalization 
method for a finite number of sites yields a set of gapless points which are 
in good agreement with the N\'eel state calculations for all values of the 
dimerization, but the agreement with the tunneling values is not very good 
even for large $S$. This raises questions about possible corrections to the 
tunneling results.
\end{abstract}
\maketitle
\vskip .6 true cm

\section{\bf I. Introduction}

One-dimensional quantum spin systems have been studied extensively for many 
years, particularly after Haldane predicted theoretically that Heisenberg 
integer spin chains should have a gap between the ground state and the first 
excited state \cite{haldane}, and this was then observed experimentally in a 
spin-1 system \cite{buyers}. Haldane's analysis used a non-linear 
$\sigma$-model (NLSM) which is a field theoretic description of the 
long-distance and low-energy modes of the spin system 
\cite{affleck1,fradkin,auerbach,sierra,naveen}.

Although the NLSM approach is supposed to be accurate only for large values
of the spin $S$, it is found to be qualitatively correct even for small values
of $S$. For instance, if there is a dimerization in the nearest 
neighbor Heisenberg couplings, taken to be 1 and $\ka$ alternately, the NLSM 
predicts that there is a discrete set of values of $\ka$ lying in the range 
$0 \le \ka \le 1$ for which the spin chain is gapless; these correspond
to quantum phase transitions. Further, the
number of gapless points is predicted to be the number of integers $\le S + 
1/2$; in particular, the undimerized chain (with $\ka =1$) is a gapless point
if $S$ is a half-odd-integer. Numerical analysis shows this to be true for 
values of $S$ up to 2 \cite{kato,pati,kitazawa1,nakamura}; however, the 
numerically obtained values of $\ka$ at the gapless points do not agree well 
with the NLSM values. It therefore appears that there must be corrections to 
the NLSM analysis for small values of $S$. 

The NLSM approach that has been used so far to find the gapless points is 
based on certain properties of a field theory in two dimensions (one space and
one time); the gapless points occur when the coefficient of a topological term
is given by $\pi$ modulo $2 \pi$ (as will be discussed in Sec. II). Although 
there are arguments justifying this criterion \cite{shankar}, there does not 
seem to be a simple physical picture behind it. One of the aims of our paper 
will be to provide a picture based on tunneling between two classical
ground states.

Numerically, there are different ways of finding the gapless points for a 
dimerized spin chain. One of the most accurate ways is based on exact 
diagonalization studies of a finite spin chain with a twisted boundary 
condition, to be specified more precisely in Sec. II 
\cite{kitazawa1,kitazawa2}. In the presence of the twist, it is found that 
the gap between the ground state and first excited state vanishes at a value 
of $\ka$ which is a function of the number of sites $2N$ (we use this notation
since the number of sites will always be taken to be even). We will be mainly 
interested in finite system sizes in our work; however, it is known from
conformal field theory \cite{kitazawa1,kitazawa2,affleck2,okamoto,nomura} 
that the locations of gapless points for the infinite system can be found very
accurately by finding the locations of those points for finite systems, and 
then extrapolating to $N \to \infty$.

In this paper, we will use three different approaches to study finite systems 
with a twisted boundary condition, in order to find the gapless points as a 
function of $S$ and $N$. The first two approaches will be analytical; they will
be based on the idea that in the presence of a twist, the system has two 
classical ground states, called N\'eel states, which are degenerate. The 
degeneracy may be broken in quantum mechanics by tunneling. However, if the
tunneling amplitude is zero, we get a gapless point in the sense that the 
lowest two states have the same energy. The third approach will be numerical 
and will be based on exact diagonalization of finite systems using various 
symmetries.

In Sec. II, we will first define the dimerized quantum spin chain and review
how it can be described using a NLSM field theory. We will then describe the 
twisted boundary condition for a finite system. In Sec. III, we will describe 
our first analytical approach. This is based on a tunneling calculation for a 
chain with a finite number of sites. For reasons explained below, this method 
is limited to small values of $\ka$. We will see that the expressions for the 
scale of the gap and the locations of the gapless points agree with those 
obtained from the NLSM field theory; this is remarkable because our analysis 
will be based only on quantum mechanical tunneling in a finite system, while 
the field theoretic analysis is based on a renormalization group equation and 
the presence of a topological term. Our second approach, described in Sec. IV,
is based on a direct quantum mechanical calculation of the matrix element 
between the two N\'eel states to lowest order in the Hamiltonian; this method 
works for all values of $\ka$. We will see that the locations of the gapless 
points obtained by the two methods differ substantially for small $\ka$. We 
make some speculations about how the tunneling results may be corrected. In 
Sec. V, we use exact diagonalization to find the gaps and the locations of the
gapless points as functions of $\ka$ for different values of $S$ and $N$. We 
find that the numerical results for the locations of the gapless points agree 
quite well with those found by the second analytical method, and therefore 
disagree with those found by the tunneling approach. In Sec. VI, we will 
summarize our results.

\section{\bf II. Dimerized quantum spin chain} 

\subsection{A. Field theoretic description}

In this subsection, we will briefly review the NLSM field theory for a 
dimerized spin chain with an infinite number of sites. The Hamiltonian is 
given by 
\beq
H ~=~ \sum_i ~[ {\vec S}_{2i-1} \cdot {\vec S}_{2i} ~+~ \ka ~{\vec S}_{2i} 
\cdot {\vec S}_{2i+1}] ~, 
\label{ham1}
\eeq
where we have spin $S$ at every site, and $\ka \ge 0$. This describes an
Heisenberg antiferromagnetic spin chain since all the couplings are positive.
We will set $\hbar = 1$, so that ${\vec S}_i^2 = S(S+1)$. (We will specify the
appropriate boundary conditions when we discuss finite systems below). In many
papers, the nearest neighbor couplings are taken to be $1 + \de$ and $1 - \de$,
instead of 1 and $\ka$. After a re-scaling of the Hamiltonian, we see that the
two parameters are related as
\beq
\ka ~=~ \frac{1 - \de}{1 + \de} ~.
\eeq

In the classical limit $S \to \infty$, the ground state of Eq. (\ref{ham1}) is
given by a configuration in which all the spins at odd sites point in the same
direction while all the spins at even sites point in the opposite direction. 
This motivates us to define a variable
\beq
{\vec \phi} (x) = \frac{{\vec S}_{2n-1} ~-~ {\vec S}_{2n}}{2S} ~,
\label{phi}
\eeq 
where $x=2na$ denotes the spatial coordinate and $a$ is the lattice spacing; 
$x$ becomes a continuous variable in the limit $a \to 0$. In the 
classical limit, $\vec \phi$ becomes a unit vector in three dimensions; the 
model is called the NLSM because of this non-linear constraint. One can 
then derive an action in terms of the field ${\vec \phi} (x,t)$; this takes 
the form \cite{haldane,affleck1}
\bea
{\cal S} &=& \int ~dt dx ~[~ \frac{1}{2cg} (\frac{\pa {\vec \phi}}{\pa t} )^2 ~
-~ \frac{c}{2g} ( \frac{\pa {\vec \phi}}{\pa x} )^2 ~]~ \non \\
& & + ~\frac{\theta}{4 \pi} ~\int ~dt dx ~{\vec \phi} ~\cdot ~
\frac{\pa {\vec \phi}}{\pa t} ~\times ~\frac{\pa {\vec \phi}}{\pa x} ~,
\label{action}
\eea
where
\bea
c &=& 2 a S ~\sqrt{\ka} ~, \non \\
g &=& \frac{1}{S} ~\frac{1 + \ka}{\sqrt{\ka}} ~, \non \\
{\rm and} \quad \theta &=& 4 \pi S ~\frac{\ka}{1 + \ka} ~.
\label{para}
\eea
The parameters $c$, $g$ and $\theta$ denote the spin wave velocity, the 
strength of the interactions between the spin waves (even though the first two
terms in (\ref{action}) are quadratic, it describes an interacting theory 
because of the non-linear constraint on $\vec \phi$), and the coefficient of a
topological term respectively. One can show that the term multiplying $\theta$
in (\ref{action}) is topological in the sense that its value is always an 
integer.

It is known that the system governed by Eq. (\ref{action}) is gapless if 
$\theta = \pi$ modulo $2 \pi$ and $g$ is less than a critical value
\cite{haldane,affleck1,shankar}. This 
implies that the theory is gapless if $4S \ka/(1+\ka) = 1, 3, \cdots$ and $g$ 
is small enough. In particular, this means that in the range $0 \le \ka \le 1$,
there are a discrete set of values of $\ka$ for which the system is gapless; 
the number of such values is given by the number of integers $\le S +1/2$. For
all values of $\theta \ne \pi$ modulo $2 \pi$, the system is gapped. For 
$\theta = 0$ modulo $2 \pi$, the gap is given by $\exp (-2 \pi /g)$. This
follows from the fact that the interaction $g$ effectively becomes a function
of the length scale $L$ and satisfies a renormalization group equation of 
the form $dg_{eff} /d \ln L = g_{eff}^2 /(2\pi)$. This implies that 
$g_{eff} (L)$ becomes very large at a length scale given by $L_0 \sim ~a
\exp (2\pi /g)$, where $g = g_{eff} (a)$ is given in Eq. (\ref{para}). This 
is the correlation length of the system; the energy gap is related to the 
inverse of this length, namely, $\Delta E \sim ~c \exp (-2 \pi /g)$.

\subsection{B. Twisted boundary condition}

In this subsection, we will study the same model as in Eq. (\ref{ham1}) but
with a finite number of sites going from 1 to $2N$. Although a periodic 
boundary condition would appear to be the simplest, it turns out that a more 
useful boundary condition is one with a twist \cite{kitazawa1,kitazawa2,kolb}.
We define the Hamiltonian to be
\bea
H &=& \sum_{n=1}^N ~[ S_{2n-1}^x S_{2n}^x + S_{2n-1}^y S_{2n}^y + S_{2n-1}^z 
S_{2n}^z] \non \\
&+& \ka ~\sum_{n=1}^{N-1} ~[ S_{2n}^x S_{2n+1}^x + S_{2n}^y S_{2n+1}^y + 
S_{2n}^z S_{2n+1}^z] \non \\
&+& \ka ~[ - S_{2N}^x S_1^x - S_{2N}^y S_1^y + S_{2N}^z S_1^z] ~.
\label{ham2}
\eea
Note that the bond going from site $2N$ to site $1$, to be called the bond
$(2N,1)$ for short, has a minus sign for the $xx$ and $yy$ couplings. We will 
call this a twisted boundary condition; it is equivalent to rotating the $x$ 
and $y$ components of ${\vec S}_1$ by $\pi$ about the $z$ axis just for that
bond.

An advantage of the twisted boundary condition is that classically, the
Hamiltonian in (\ref{ham2}) has exactly two ground states, namely,
(i) ${\vec S}_{2n-1} = (0,0,S)$ and ${\vec S}_{2n} = (0,0,-S)$ for all $n$, 
and (ii) ${\vec S}_{2n-1} = (0,0,-S)$ and ${\vec S}_{2n} = (0,0,S)$ for all
$n$. We will call these two N\'eel states $N_1$ and $N_2$ respectively.
(This is in contrast to the case of periodic boundary conditions where there 
is an infinite family of classical ground states because ${\vec S}_{2n-1} = 
- {\vec S}_{2n}$ can point in any direction). If the classical degeneracy is 
broken quantum mechanically in any way, there will be a gap between the lowest
two states of the system, while if the degeneracy remains unbroken, the system
will be gapless. In Secs. III and IV, we will describe two ways of analytically
studying whether the degeneracy is broken. 

Let us now describe the various symmetries of the Hamiltonian in Eq. 
(\ref{ham2}). Although the total spin is not a good quantum number because of 
the twist on one bond, $S_{tot}^z = \sum_n (S_{2n-1}^z + S_{2n}^z)$ is a good 
quantum number.

Eq. (\ref{ham2}) satisfies the duality property $H (\ka) = \ka {\tilde H} 
(1/\ka)$, where $\tilde H$ is related to $H$ by an unitary transformation. 
[The unitary transformation is required because the twist only exists at the 
bond $(2N,1)$ whose strength is $\ka$. By applying a rotation $S_1^x \to - 
S_1^x$ and $S_1^y \to - S_1^y$, we can move the twist from the bond $(2N,1)$ 
to the bond $(1,2)$.] Since a unitary transformation does not affect the 
spectrum, we conclude that if there is a gapless point at a value $\ka$, 
there must also be a gapless point at $1/\ka$.

Next, we define the parity transformation $P$ as reflection of the system
about the bond $(2N,1)$, namely, 
\beq 
{\vec S}_i \leftrightarrow {\vec S}_{2N+1-i} ~. 
\label{parity}
\eeq
This is a discrete symmetry of the Hamiltonian $H$, and all eigenstates of $H$
will be eigenstates of $P$ with eigenvalue $\pm 1$. For any value of $\ka$,
we find that the ground and first excited states always have opposite values 
of the parity. We will show below that the relative parity of the ground 
states in the limits $\ka \to 0$ and $\ka \to \infty$ is given by
$(-1)^{2S}$. For integer $S$, this will imply that in the range $0 < \ka < 
\infty$, the number of crossings between the ground state and the first 
excited state (and hence the number of gapless points) must be even. But for 
half-odd-integer $S$, the number of such crossings must be odd; combined with 
the duality $\ka \to 1/\ka$, this implies that there must be a crossing
and therefore a gapless point at $\ka = 1$ (this is a self-dual point).

To prove the statement about the relative parities of the ground states at 
$\ka \to 0$ and $\ka \to \infty$ being given by $(-1)^{2S}$, we proceed as 
follows. We first observe that if there are only two spins 1 and 2 governed 
by the untwisted Hamiltonian $h_u = S_1^x S_2^x + S_1^y S_2^y + S_1^z 
S_2^z$, then under the reflection $1 \leftrightarrow 2$, the ground state 
has the parity $(-1)^{2S}$, while the first excited state has the parity 
$(-1)^{2S+1}$. But for the twisted Hamiltonian $h_t = - S_1^x S_2^x - 
S_1^y S_2^y + S_1^z S_2^z$, the ground state and first excited
state have parities equal to $1$ and $-1$ respectively. (This can be proved
using the Perron-Frobenius theorem for a real symmetric matrix). We now
consider the entire system with $2N$ sites. In the limit $\ka \to 0$,
the ground state of the system is given by a direct product of ground states
over the bonds $(1,2)$, $(3,4)$, ..., $(2N-1,2N)$. Since there are $N$ dimers,
the parity of this state is $(-1)^{2SN}$, while the parity of the first 
excited state is $(-1)^{2SN+1}$. On the other hand, in the limit $\ka \to 
\infty$, the ground state of the system is given by a direct product of ground
states over the bonds $(2,3)$, $(4,5)$, ..., $(2N,1)$. Under parity, 
the parity of this state is $(-1)^{2SN+2S}$, while the parity of the first 
excited state is $(-1)^{2SN+2S+1}$. A comparison between the ground states
in the two limits shows that they have a relative parity of $(-1)^{2S}$.

\section{\bf III. Tunneling approach to the finite spin chain}

In this section, we will study the model defined in Eq. (\ref{ham2}) using a
tunneling approach. We are interested in the limit $S \to \infty$ and $\ka \to
0$, such that $\ka S$ is of order 1. We will compute the action of the system 
and use that to compute the tunneling amplitude between the two N\'eel states.

In the limit $\ka \to 0$, the system consists of decoupled dimers whose energy
levels are easy to compute. (We will assume in this section that there are at 
least three dimers, i.e, $N \ge 3$). For the dimer on the bond $(2n-1,2n)$, we 
define the variables
\bea
{\vec \phi}_n &=& \frac{{\vec S}_{2n-1} ~-~ {\vec S}_{2n}}{2S} ~, 
\non \\
{\vec l}_n &=& {\vec S}_{2n-1} ~+~ {\vec S}_{2n} ~. 
\label{phil1}
\eea
These variables satisfy the relations
\bea
{\vec \phi}_n^2 &=& 1 ~+~ \frac{1}{S} ~-~ \frac{{\vec l}_n^2}{4S^2} ~, \non \\
{\vec \phi}_n ~\cdot ~{\vec l}_n &=& 0 ~, \non \\
\left[ l_m^a ~,~ \phi_n^b \right] &=& i ~\de_{mn} ~\sum_c ~\ep^{abc} ~
\phi_n^c ~, \non \\
\left[ l_m^a ~,~ l_n^b \right] &=& i ~\de_{mn} ~\sum_c ~\ep^{abc} ~l_n^c ~, 
\non \\
{\rm and} \quad \left[ \phi_m^a ~,~ \phi_n^b \right] &=& \frac{i}{4S^2} ~
\de_{mn} ~\sum_c ~\ep^{abc} ~l_n^c ~.
\label{phil2}
\eea

Several simplifications occur in the classical limit $S \to \infty$. Firstly, 
in the N\'eel state, $\vec l$ is equal to zero while $\vec \phi$ is an unit 
vector; the latter is clear from the first equation in (\ref{phil2}). We will 
therefore set ${\vec \phi}_n^2 = 1$ exactly. Secondly, we will take the 
commutator $[ \phi_m^a , \phi_n^b ] = 0$ due to the fourth equation in 
(\ref{phil2}). Finally, given the second and third equations in (\ref{phil2}),
we can obtain the momentum which is canonically conjugate to $\phi_n$, namely,
${\vec l}_n = {\vec \phi}_n ~\times ~ {\vec \Pi}_n$, which satisfies the 
commutation relation
\beq
\left[ \phi_m^a ~,~ \Pi_n^b \right] ~=~ i ~\de_{mn} ~\de_{ab} ~.
\eeq

Using Eq. (\ref{phil1}), we can write the Hamiltonian in (\ref{ham2}) in
terms of $\vec \phi$ and $\vec \Pi$, and then obtain the Lagrangian as
\beq
L ~=~ \sum_{n=1}^N ~\frac{d {\vec \phi}_n}{d t} ~\cdot ~
{\vec \Pi}_n ~-~ H ~.
\eeq
We eventually find that
\bea
L &=& \sum_{n=1}^N ~\frac{1}{2} ~( \frac{d {\vec \phi}_n}{d t} )^2 ~
+~ \ka S^2 ~\sum_{n=1}^{N-1} ~{\vec \phi}_n \cdot {\vec \phi}_{n+1} 
\non \\
& & +~ \ka S^2 [~- \phi_N^x \phi_1^x ~-~ \phi_N^y \phi_1^y ~+~ \phi_N^z 
\phi_1^z ~] \non \\
& & + ~\frac{\ka S}{2} ~\sum_{n=1}^{N-1} ~(\frac{d{\vec \phi}_n}{dt} ~+~ 
\frac{d{\vec \phi}_{n+1}}{dt}) \cdot {\vec \phi}_n \times {\vec \phi}_{n+1} 
\non \\
& & + ~\frac{\ka S}{2} ~[~~ (\frac{d \phi_N^x}{dt} ~-~ \frac{d \phi_1^x}{dt})
(\phi_N^y \phi_1^z ~+~ \phi_N^z \phi_1^y) \non \\
& & ~~~~~~~ + ~(\frac{d \phi_N^y}{dt} ~-~ \frac{d \phi_1^y}{dt})
(-\phi_N^z \phi_1^x ~-~ \phi_N^x \phi_1^z) \non \\
& & ~~~~~~~ + ~(\frac{d \phi_N^z}{dt} ~+~ \frac{d \phi_1^z}{dt})
(-\phi_N^x \phi_1^y ~+~ \phi_N^y \phi_1^x) ~] \non \\
& & +~ \ka^2 S^2 ~[ {\rm terms ~of ~fourth ~order ~in ~{\vec \phi}}_n ]~.
\label{lag}
\eea
The terms in the third line of this Lagrangian are what give rise to the 
topological term in Eq. (\ref{action}) in the continuum limit. In the last 
line of (\ref{lag}), the terms of fourth order in ${\vec \phi}_n$ are chosen 
in such a way that when we compute the Hamiltonian from it, it agrees with Eq.
(\ref{ham2}). We will now see why these fourth order terms are not important. 

In the limit $\ka \to 0$, $S \to \infty$ and $\ka S$ of order 1, we can scale 
the time $t$ by a factor of $\sqrt S$ to show that only the first two lines of
Eq. (\ref{lag}) contribute to the Euler-Lagrange equations of motion (EOM); 
this is a major simplification. To compute the tunneling amplitude between the
two N\'eel states, we will find the solutions of the EOM in imaginary time. 
We then find that the tunneling amplitude comes with a phase which arises from
the third through sixth lines of Eq. (\ref{lag}); thus these terms are 
important even though they do not directly contribute to the EOM. The fourth 
order terms in the last line of (\ref{lag}) do not contribute to either the 
EOM or the phase, and we will therefore ignore them henceforth. 

In imaginary time (denoted by the symbol $\tau$), the action takes the form 
\bea
{\cal S}_I &=& \int d\tau ~[ \sum_{n=1}^N ~\frac{1}{2} ~( \frac{d 
{\vec \phi}_n}{d \tau} )^2 + \ka S^2 (N - \sum_{n=1}^{N-1} ~{\vec \phi}_n
\cdot {\vec \phi}_{n+1}) \non \\
& & ~~~~~~~ +~ \ka S^2 [~ \phi_N^x \phi_1^x ~+~ \phi_N^y \phi_1^y ~-~ \phi_N^z
\phi_1^z ~] \non \\
& & -i ~\frac{\ka S}{2} ~\sum_{n=1}^{N-1} ~(\frac{d{\vec \phi}_n}{d\tau} ~+~ 
\frac{d{\vec \phi}_{n+1}}{d\tau}) \cdot {\vec \phi}_n \times {\vec \phi}_{n+1}
\non \\
& & -i ~\frac{\ka S}{2} ~[~~ (\frac{d \phi_N^x}{d\tau} ~-~ \frac{d 
\phi_1^x}{d\tau}) (\phi_N^y \phi_1^z ~+~ \phi_N^z \phi_1^y) \non \\
& & ~~~~~~~ + ~(\frac{d \phi_N^y}{d\tau} ~-~ \frac{d \phi_1^y}{d\tau})
(-\phi_N^z \phi_1^x ~-~ \phi_N^x \phi_1^z) \non \\
& & ~~~~~~~ + ~(\frac{d \phi_N^z}{d\tau} ~+~ \frac{d \phi_1^z}{d\tau})
(-\phi_N^x \phi_1^y ~+~ \phi_N^y \phi_1^x) ~] ~] ~. \non \\
& &
\label{si}
\eea
(We have introduced a constant $\ka S^2 N$ so that the action vanishes for
each of the two N\'eel states). The tunneling amplitude will be given by the 
sum of $\exp (-{\cal S}_I)$ along all the paths of extremal action which join 
the N\'eel states. We will now determine these extremal paths.

Let us use polar angles to write the variables ${\vec \phi}_n = 
(\sin \al_n \cos \be_n, \sin \al_n \sin \be_n , \cos \al_n)$. The N\'eel 
states 1 and 2 are given by $\al_n = 0$ for all $n$ and $\al_n = \pi$ for 
all $n$ respectively. We now solve the EOM following from the first two lines 
of Eq. (\ref{si}) in order to obtain the paths going from state 1 to state 2. 
We will not write the EOM explicitly here, but directly present the solutions.
We find that the two paths which have the least action are given by 

\noi A: ~$\al_n (\tau) = \al (\tau)$ and $\be_n = \be_0 + (n \pi /N)$ for 
all $n$, and 

\noi B: ~$\al_n (\tau) = \al (\tau)$ and $\be_n = \be_0 - (n \pi /N)$ for 
all $n$, 

\noi where $\be_0$ is an arbitrary angle.
In both cases, the function $\al (\tau)$ satisfies the EOM
\beq
\frac{d^2 \al}{dt^2} ~=~ \ka S^2 ~\sin (2 \al) ~(1 ~-~ \cos \frac{\pi}{N}) ~,
\eeq
with the boundary conditions $\al (-\infty) =0$, $\al (\infty) =\pi$, and 
$d \al (\pm \infty) /dt = 0$. This implies that 
\beq
\frac{d\al}{dt} ~=~ 2 S \sqrt{\ka} ~\sin \al ~\sin \frac{\pi}{2N} ~.
\eeq
Using this we can evaluate the contribution of the first two lines of 
(\ref{si}) along either one of the paths joining the N\'eel states. We 
find that the real part of the action is given by 
\bea
{\rm Re} ~{\cal S}_I &=& N \int_{-\infty}^\infty d\tau ~[\frac{1}{2} ~
(\frac{d\al}{d\tau})^2 + \ka S^2 \sin^2 \al ~(1 - \cos \frac{\pi}{N})] \non \\
&=& 4 \sqrt{\ka} S N ~\sin \frac{\pi}{2N} ~.
\label{real}
\eea

We can now evaluate the contribution of the imaginary terms in Eq. (\ref{si})
to the action. We find that for path A, they contribute $-i2\ka S N \sin(\pi 
/N)$, while for path B, they contribute $i2\ka S N \sin(\pi /N)$. Hence, the 
total contribution of the two paths to $\exp (-{\cal S}_I)$ is given by
\beq
\Delta ~\sim~ \cos (2\ka S N \sin \frac{\pi}{N}) ~\exp (-4 \sqrt{\ka} S N ~
\sin \frac{\pi}{2N}) ~,
\label{tungap}
\eeq
up to some pre-factors which are determined by fluctuations about the 
classical paths. Since $\Delta$ is the matrix element between two classically
degenerate states, the energy gap between the two states is given by $2 
|\Delta|$. We thus see that the gap vanishes if $2\ka S N \sin (\pi /N)$ 
is an odd multiple of $\pi /2$, i.e., if
\beq
4\ka S N \sin \frac{\pi}{N} ~=~ \pi {\rm ~modulo~} 2 \pi ~.
\label{gapless}
\eeq
This is the same condition as the one satisfied by the parameter $\theta$ in 
Sec. II A in the limits $S, N \to \infty$ and $\ka \to 0$. Further, if $4\ka 
S N \sin (\pi /N) = 0$ modulo $2 \pi$, the gap is given by $\exp (-{\rm 
Re} ~ {\cal S}_I)$ which agrees with the expression $\exp (-2\pi /g)$ given in
Sec. II A for $S, N \to \infty$ and $\ka \to 0$. Thus, a simple quantum 
mechanical tunneling calculation seems to reproduce the same conditions as 
those obtained earlier by more complex field theoretic calculations involving
topological terms and a renormalization group analysis.

Before ending this section, we should note that there are other pairs of paths
with extremal action, which have the form

\noi A: ~$\al_n (\tau) = \al (\tau)$ and $\be_n = \be_0 + (pn \pi /N)$ for 
all $n$, and 

\noi B: ~$\al_n (\tau) = \al (\tau)$ and $\be_n = \be_0 - (pn \pi /N)$ for 
all $n$, 

\noi where $p = 3, 5, \cdots$ (going up to the largest odd integer $\le N-1$)
labels the different pairs of paths. However, the real part of the action of 
these paths is given by $4 \sqrt{\ka} S N ~\sin (p\pi /2N)$, which, for large 
$S$, is much larger than the expression given in Eq. (\ref{real}); their 
contributions to the tunneling amplitude are therefore much smaller.

Finally, we would like to note that it is important that the twist
in the boundary condition should be by $\pi$, and not by any other angle.
Even though any non-zero twist would lead to two N\'eel ground states
classically, the pairs of tunneling paths between those two ground states
would not have the same real part of the action if the twist angle was 
different from $\pi$. The pairs of paths would therefore not cancel each
other no matter what the imaginary parts of their actions are. (Two complex 
numbers cannot add up to zero, no matter what their phases are, if their 
magnitudes are not equal).

\section{\bf IV. A second approach to the matrix element between N\'eel 
states}

In the previous section we argued that the gapless points of the Hamiltonian 
in Eq. (\ref{ham2}) can be identified with the values of $\ka$ for which 
the tunneling amplitude between the two classical ground states vanish. In 
this section we will calculate this amplitude using an alternate method. 

The twist on the edge bond $(2N,1)$ breaks the global $SU(2)$ symmetry and thus
lifts the continuous degeneracy of the classical ground states. With the twist,
the two degenerate ground states of the Hamiltonian are the N\'eel states 
which are connected to each other by rotation by $\pi$ about the $y$ axis,
\bea
| N_1 \rangle &=& |S, -S, S, \cdots, -S, S, -S \rangle ~, \non \\
{\rm and} \quad | N_2 \rangle &=& |-S, S, -S, \cdots, S, -S, S \rangle ~,
\label{neel1}
\eea
where $| \{m_i\} \rangle $ denotes the state with $S^z_i$ eigenvalue $m_i$. 

We are interested in the zeroes of the quantity
\beq
T = \langle N_2 | e^{-\beta H} |N_1 \rangle ~,
\label{transamp}
\eeq
as a function of $\ka$. Though the calculation of the above matrix element 
is an interacting many-body problem, we can obtain its zeroes `exactly' in 
the thermodynamic limit. 

First we note that, in the expansion of the exponential,
\beq
e^{-\beta H} = \sum_{n=0}^\infty ~\frac{(-\beta H)^n}{n!}~,
\eeq
the first term which makes a non-zero contribution to $T$ has $n=2SN$. This 
is because to take $|N_1 \rangle$ to $|N_2 \rangle$, spins belonging to the 
$A$-sublattice have to be flipped from $|S \rangle$ to $|-S\rangle$, and 
this requires the action of $(S^-)^{2S}$ for each spin. Similarly, the action 
of $(S^+)^{2S}$ will take spins in the $B$-sublattice from $|-S\rangle$ to 
$|S\rangle$. 

Next, we will calculate the values of $\ka$ for which
\beq
\langle N_2 | H^{2SN} |N_1 \rangle = 0 ~.
\label{leadmat}
\eeq
Then we will show that as $N \to \infty$, Eq. (\ref{leadmat}) implies 
that
\beq
\langle N_2 | H^{2SN+k} |N_1 \rangle = 0
\label{subleadmat}
\eeq
for any finite $k$. This in turn will imply that $T$ is zero.

The only term in $H^{2SN}$ which makes a non-zero contribution to the left 
hand side of Eq. (\ref{leadmat}) is
\[ \prod_{i=1}^N (S_{2i}^-)^{2S}(S_{2i+1}^+)^{2S}. \]
We need to count the number of ways in which such a term can arise. The 
contribution comes from terms of the following type,
\beq
\prod_{i=1}^N (S_{2i}^-~S_{2i+1}^+)^{m}(S_{2i-1}^+ ~ S_{2i}^-)^{2S-m} ~,
\label{nonzero}
\eeq
where $0 \le m \le 2S$. The above term can be obtained in
\[\frac{(2SN)!}{(m!)^N \big((2S-m)!\big)^N} \] ways and comes with a
weight $(-1)^m \ka^{mN}$. Here we have neglected an overall
$m$-independent factor due to the Clebsch-Gordon coefficients arising
from the repeated application of $S^+$ and $S^-$ operators. The
condition in Eq. (\ref{leadmat}) then becomes,
\bea
\sum_{m=0}^{2S} ~(-1)^m ~\Big(a_m(\ka) \Big)^N &=& 0 ~, \non \\
{\rm where} \quad a_m(\ka) &=& \frac{\ka^m}{m!(2S-m)!} ~.
\label{roots}
\eea
Before proceeding further, we note that the above condition preserves the 
duality symmetry under $\ka \to 1/\ka$. This is because 
Eq. (\ref{roots}) can be written as,
\beq
\ka^{2SN} (-1)^{2S} \sum_m (-1)^m \Big(a_m(1 / \ka)\Big)^N = 0 ~.
\label{dualroots}
\eeq
Hence, if $\ka^*$ is a solution, so is $1/\ka^*$. Thus we can restrict 
ourselves to the range $0 \le \ka \le 1$.

Eq. (\ref{roots}) determines the roots of a polynomial of order $2SN$, which, 
in general, cannot be solved analytically. But it turns out that we can obtain
the roots in the limit $N \to \infty$. In this limit, depending on the value 
of $\ka$, one particular term in the sum is predominant, and the rest of the 
terms can be neglected compared to this. The dominant term is determined by,
\beq
\max_m ~a_m \equiv a_{m^*} ~.
\label{dominant}
\eeq
As $\ka$ varies from $0$ to $1$, $m^*$ successively takes values
$m^* = 0, 1, \cdots , S$ for integer $S$ and $m^* = 0, 1,
\cdots , S+1/2$ for half-odd-integer $S$. Noting that neighboring terms in 
the sum have opposite signs, Eq. (\ref{roots}) can be satisfied only when
\beq
a_m = a_{m+1} ~.
\label{roots1}
\eeq
Thus the gapless points are given by
\beq
\ka^*_m = \frac{m+1}{2S-m} ~,
\label{gaplesspts}
\eeq
where, $m = 0, 1, \cdots , S-1$ for integer $S$ and $m = 0, 1, \cdots , S-1/2$
for half-odd-integer $S$.

To complete our argument that Eq. (\ref{gaplesspts}) gives the gapless points,
we need to show that Eq. (\ref{leadmat}) implies Eq. (\ref{subleadmat}). To 
this end, we first note that the non-zero contributions to the matrix element 
from $H^{2SN+k}$ can be obtained from the contributing terms in $H^{2SN}$, 
given in Eq. (\ref{nonzero}), by adding terms of the form $S_i^z S_{i+1}^z$, 
$S_i^+ S_{i+1}^-$ or $S_i^- S_{i+1}^+$. Now the weight coming from the
Clebsch-Gordon coefficients depend on the order in which the terms appear. 
But formally one can write that, Eq. (\ref{subleadmat}) implies
\beq
\sum_m (-1)^m b_{m,k} \Big(a_m(\ka) \Big)^N = 0 ~, 
\label{nonzerok}
\eeq
where $b_{m,k}$ are finite undetermined constants independent of $N$. As 
before, in the large-$N$ limit, the left hand side of Eq. (\ref{nonzerok})
can be zero only through the mutual cancellation of a pair of neighboring
terms, {\it i.e.}, when
\beq
\frac{a_m}{a_{m+1}} = \left( \frac{b_{m+1}}{b_{m}} \right)^{1/N} ~.
\eeq
As $N \to \infty$, this reduces to the condition in Eq. (\ref{roots1}).
In other words, the vanishing of $\langle N_2 | H^{2SN} | N_1 \rangle$ is a 
sufficient condition for the vanishing of $\langle N_2 | H^{2SN+k} | N_1 
\rangle$ for any finite $k$ as $N \to \infty$.

For half-odd-integer spins, $\ka^* = 1$ is a solution of Eq. (\ref{gaplesspts})
for $m = S-1/2$, but it is {\it not} a solution for integer spins. This is 
consistent with Haldane's conjecture \cite{haldane} that the uniform chain is 
gapless for half-odd-integer spins and gapped for integer spins.

Since the identification of gapless points with the zeroes of the transition 
amplitudes between the two N\'eel states is essentially a semi-classical 
approximation, we expect the formula given by Eq. (\ref{gaplesspts}) to get 
better for larger values of $S$.

As with the tunneling calculation in Sec III, here also one can see that it 
is crucial to have a twist by $\pi$ and not any other angle. Let us suppose 
that the twist angle is $\chi$. Then the Hamiltonian for the bond between the 
spins at sites $2N$ and $1$ will be
\[ \ka (S_{2N}^z S_1^z + e^{i\chi}~ S_{2N}^+ S_1^- + e^{-i\chi}~ 
S_{2N}^- S_1^+). \]  
Then the equivalent of the condition in Eq. (\ref{roots1}) will read
\beq
a_m = -e^{i\chi} a_{m+1}.
\label{roots2}
\eeq
Since $a_m$'s are all real and positive, such a condition can be satisfied only when
$\chi = \pi$, in which case Eq. (\ref{roots2}) becomes
Eq. (\ref{roots1}). 

Finally, let us compare the gapless values of $\ka$ given in Eq. 
(\ref{gaplesspts}) with those given in Eq. (\ref{gapless}) in the limit
$N \to \infty$, namely,
\beq
\ka^*_m = \frac{m+1/2}{2S} ~,
\label{gapless2}
\eeq
where, $m = 0, 1, \cdots$. [Since Eq. (\ref{gapless}) was derived under the
assumption that $S \to \infty$ and $\ka S$ is of order 1, we must restrict $m$
to be much less than $S$ in Eq. (\ref{gapless2}).] We see that for large 
values of $S$ and $m \ll S$, the values of $\ka^*_m$ in Eqs. (\ref{gaplesspts})
and (\ref{gapless2}) are related by a shift of $1/(4S)$. It
would be useful to understand more deeply why this is so. Heuristically, this 
discrepancy can be explained by postulating that the phase difference between 
the actions for the two paths discussed in Sec. III has an additional factor 
of $\pi$ for some reason. Eq. (\ref{gapless}) would then change to $4\pi \ka 
S = 0$ modulo $2 \pi$ for $N \to \infty$; this condition would be equivalent 
to Eq. (\ref{gaplesspts}) for $m \ll S$. In a different problem (tunneling
of a charged particle in two dimensions in the presence of a large magnetic 
field), it was empirically found that an additional factor of $\pi$ appears 
due to fluctuations about the tunneling paths \cite{jain}. It may 
be worth studying if something similar happens in our problem.
 
\section{V. Numerical results for finite systems}

In this section, we numerically determine the values of $\ka$ for which the 
Hamiltonian becomes gapless using exact diagonalization
of finite systems. These results, being a direct calculation of the gapless
points, will give us information about the physical regimes in which the 
analytical methods outlined in Secs. III and IV are valid.

We begin with a brief outline of the method used to find the gapless points.
As we vary $\ka$ in Eq. (\ref{ham1}), it is known that the gapless points 
separate various phases whose ground states are represented approximately by 
different valence bond states (see Refs. \cite{kitazawa1,oshikawa} and 
references therein). Specifically, for a given spin $S$, there are $2S+1$ 
different phases separated by the $2S$ gapless points for $\ka$ between $0$ 
and $\infty$. The lowest two eigenstates of an untwisted Hamiltonian never 
cross each other in energy, even at the transition from one phase to the other;
also, the ground state always has the same eigenvalue $(-1)^{2SN}$ for the 
parity $P$. It is here that we make use of the twisted boundary condition in 
Eq. (\ref{ham2}). It has been shown \cite{kitazawa1} that the lowest two 
eigenstates of this Hamiltonian (both of which lie in the $S_{tot}^z=0$ sector)
have different parity eigenvalues, and they cross at certain points which, in 
the limit $N \to \infty$, become the gapless points of the Hamiltonian in 
Eq. (\ref{ham1}). This means
that one can locate the gapless points of the Hamiltonian in Eq. (\ref{ham1}) 
by studying the crossing of the two lowest eigenvalues of the Hamiltonian in 
Eq. (\ref{ham2}) in {\it different parity sectors}. This enables us to 
find the gapless points without having to consider two very closely spaced 
eigenvalues lying within a single symmetry sector.

We study the Hamiltonian in Eq. (\ref{ham2}) numerically. Because of the 
duality between $\ka$ and $1/\ka$ we restrict our studies to $0~\leq~\ka~
\leq 1$. We use the Lanczos algorithm to diagonalize finite systems from $N=3$
to $N=5$. Spins from $S=2$ to 3 are studied for $N=3$ to $N=5$, but for 
$S = 3.5$ to 7 we restrict ourselves to $N=3$ and $N=4$ due to computational
limitations.

\begin{figure}[htb]
\begin{center}
\epsfig{figure=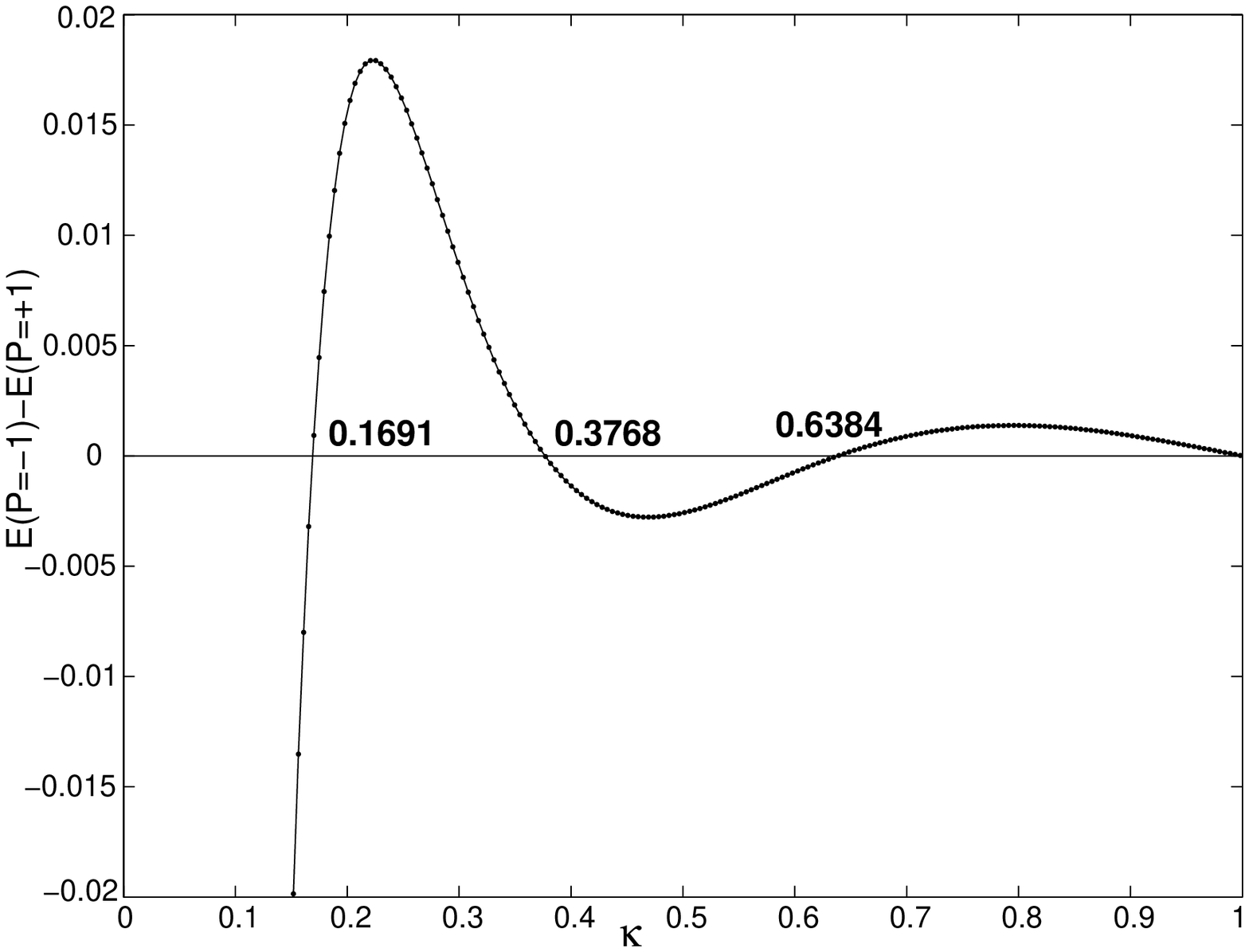,width=8cm}
\vskip .2 true cm
\epsfig{figure=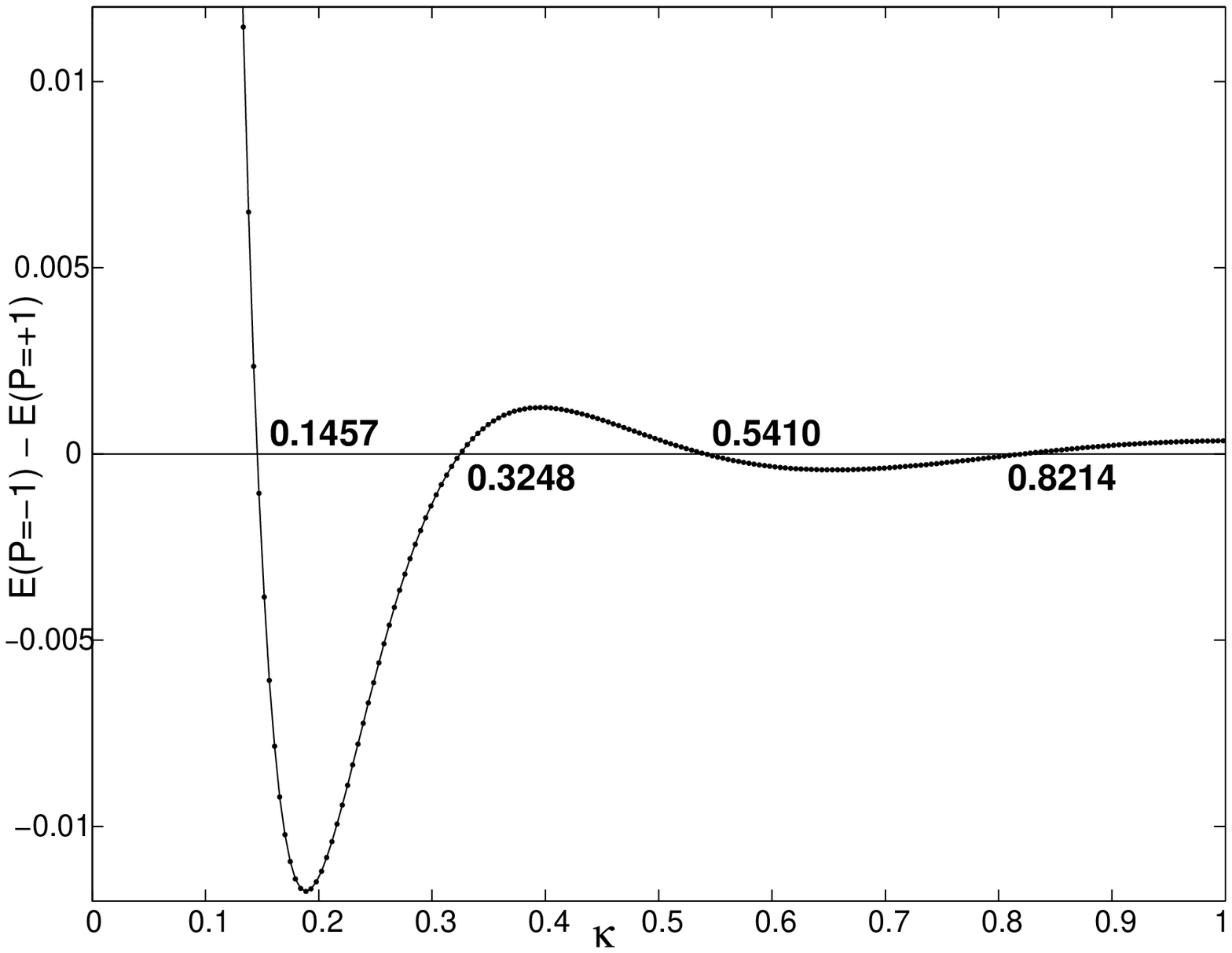,width=8cm}
\end{center}
\caption{Energy difference between the two lowest energies in the $P=-1$ and 
$P=+1$ sectors as a function of $\ka$, for $N=3$. The upper figure is for 
$S=3.5$, and the lower figure is for $S=4$. The locations of the gapless
points are shown.}
\end{figure}

In Fig. 1, we present two representative data plots, for $N=3$. 
The upper plot is for $S=3.5$ and the lower one is 
for $S=4$. Plotted on the $y$ axis in both figures is $E(P=-1) - E(P=+1)$, 
the energy difference between the two lowest energy eigenstates in the 
two parity sectors. Wherever the plot crosses the $x$ axis we have a gapless 
point of the Hamiltonian in Eq. (\ref{ham1}). The values of the crossing
points are indicated in the plots. From the discussion in Sec. II, we know 
that the ground state for $N=3$, $S=3.5$ should have a parity $-1$, and for 
$N=3$, $S=4$ should have a parity $+1$. This is indeed borne out by the plots 
in the figure. Moreover, the gapless point at $\ka=1$ is also present for 
$S=3.5$ as expected. The nature of these plots for other spins and different 
lattice sizes is similar. Since our emphasis in this work is on the locations
of the gapless points, we now turn to analyzing those points more closely.
[Incidentally, we observe in Fig. 1 that the envelope of the magnitude of the 
gap is rapidly decreasing with increasing $\ka$; this is in accordance with 
the exponential factor for the tunneling amplitude in Eq. (\ref{tungap}).] 

\begin{figure}[htb]
\begin{center}
\epsfig{figure=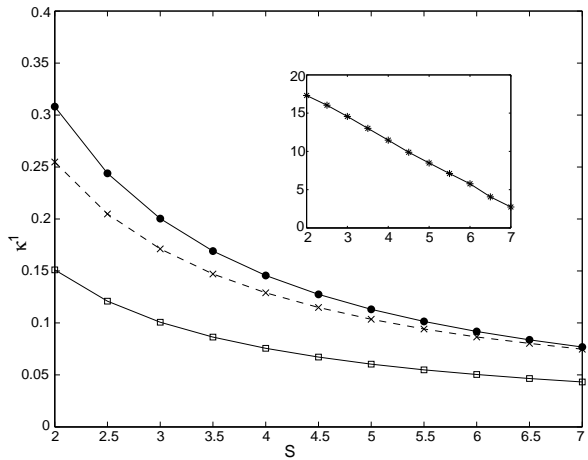,width=8cm}
\end{center}
\caption{The location $\ka^1$ of the gapless point closest to zero, as a 
function of the spin, for $N=3$. The results from numerical calculations 
(dots), N\'eel state calculations (crosses) and tunneling calculations 
(squares) are shown. The joining lines are guides for the eye. The inset shows
the percentage variation of the N\'eel state calculations when compared with 
the numerical results.}
\end{figure}

\begin{figure}[htb]
\begin{center}
\epsfig{figure=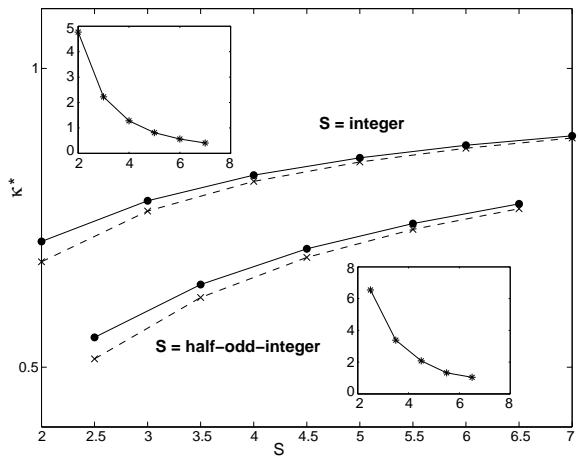,width=8cm}
\end{center}
\caption{The location $\ka^*$ of the gapless point closest to (but less than) 
1, as a function of the spin, for $N=3$. The results from numerical 
calculations (dots) and N\'eel state calculations (crosses) are shown. The
top and bottom parts of the figure are for integer and half-odd-integer
values of the spin respectivly. The joining lines are guides for the eye. The 
insets shows the percentage variation of the N\'eel state calculations when 
compared with the numerical results.}
\end{figure} 

Figure 2 shows a comparison of the different methods used to calculate the 
gapless points. Plotted on the $y$ axis is $\ka^1$, the gapless point closest 
to the origin for various values of the spin, for $N=3$. The topmost plot 
(marked by dots) is for values obtained from the numerical calculations, 
called $\ka^1_{num}$. The next plot (crosses) is for values 
obtained from the N\'eel state calculation in Eq. (\ref{roots}), 
$\ka^1_{Neel}$. The plot at the bottom (squares) is for values 
obtained from the tunneling expression in Eq. (\ref{gapless}), $\ka^1_{tun}$.
Clearly, the values of $\ka^1$ obtained using the N\'eel state calculation
are in closer agreement with the numerical values than the tunneling values.
The tunneling values do not converge with increasing $S$, even 
though one is looking at data for spin values as large as $S=7$.
On the other hand, the plot of values using 
the N\'eel state calculations converges much faster. This is made clearer 
in the plot by the inset where on the $y$ axis we have plotted the percentage 
deviation, ($(\ka^{1}_{num}-\ka^{1}_{Neel})/\ka^{1}_{num}) \times 100$,
of the values of $\ka^1_{Neel}$ from the numerically obtained values.

Figure 3 shows a comparison between the numerical results (dots)
and those obtained from the N\'eel state calculation (crosses) for 
$\ka^{*}$, the gapless point closest to $\ka=1$. [Unlike Fig. 2, we have not 
shown the tunneling results based on Eq. (\ref{gapless})) because that 
formula for the gapless points is not valid when $\ka$ is close to 1.] 
The data sets for integer and half-odd-integer spins have 
been plotted separately. The plot and the inset at the top are for integer
spins, and the ones at the bottom are for half-odd-integer spins. As before, 
we see that the agreement with the numerical results improves as we go to 
larger spins. We also see from the insets that for a given spin, the agreement
is much better near $\ka=1$ than it was near $\ka=0$. This suggests that the 
N\'eel state calculation gets better as we increase $\ka$ from 0 to 1.
This is something which is seen very clearly in the following tables.	

\begin{center}
\begin{tabular}{|c|c|c|c|c|} \hline
$\ka_{num}$ & $\ka_{Neel}$ & \% deviation& $\ka_{tun}$ & \% deviation \\ 
\hline 
0.083 & 0.080 & 3.6\% &0.047 &45.4\% \\
0.190 & 0.180 & 5.2\% & 0.140 &26.3\% \\
0.306 & 0.293 & 4.2\% &0.232 & 24.2\% \\
0.438 & 0.424 & 3.19\% &0.326 &25.6\% \\
0.773 & 0.765 & 1.03\%&0.512 &33.8\% \\
1& 1& 0\%& & \\ \hline
\end{tabular} 
\end{center}

\begin{center}
\begin{tabular}{|c|c|c|c|c|} \hline
$\ka_{num}$ & $\ka_{Neel}$ & \% deviation& $\ka_{tun}$ &
\% deviation \\ \hline
0.077 & 0.074 & 3.89\% &0.043 &44.2\% \\
0.174 & 0.166 & 4.02\% & 0.130 &25.3\% \\
0.281 & 0.269 & 4.98\% &0.252 & 23.1\% \\
0.400 & 0.387 & 3.25\% &0.302 &24.5\% \\
0.536 & 0.525 & 2.05\%&0.389 &27.4\% \\
0.695& 0.685& 1.29\%&0.475 & 31.7\% \\ 
0.887& 0.884& 0.34\%&0.561& 36.8\% \\\hline
\end{tabular}
\end{center}

\noi Table 1. Comparison of the values of all the gapless points obtained 
using the three methods of calculating them, for $S=6.5$ (top) and 7 (bottom).
The data presented is for $N=3$. 
\vspace{0.25cm}

Table 1 shows how the numerical results, the N\'eel state calculations and the
tunneling results compare for all the gapless points. The table shows the 
values of $\ka$ at which the Hamiltonian in Eq. (\ref{ham1}) is gapless for 
$S=6.5$ (top) and $S=7$ (bottom), for $N=3$. The percentage deviations as 
defined earlier are also shown for the N\'eel state 
and tunneling calculations. As conjectured after looking at Fig. 3, we 
see that the N\'eel state calculation gives successively better approximations
to the actual gapless points as we go further from the origin $\ka = 0$. We 
have shown the tunneling values for all the gapless points only for 
completion; the formula given by Eq. (\ref{gapless}) is valid only 
for $\ka S$ of order 1. We again see that these values have much 
larger percentage deviations from the values obtained from the other two 
methods, even though the values of the spins considered are quite large.

\begin{figure}[htb]
\begin{center}
\epsfig{figure=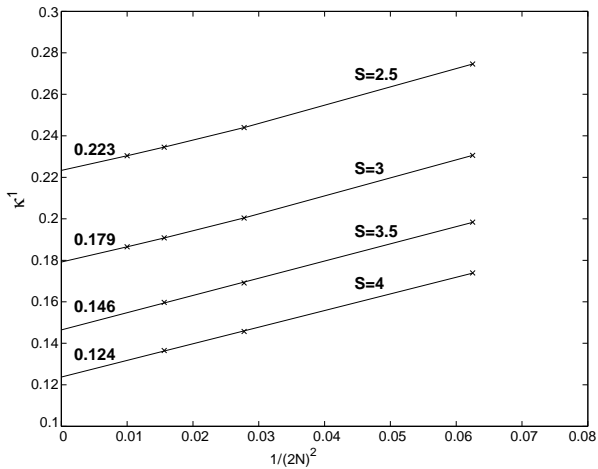,width=8cm}
\end{center}
\caption{Variation of $\ka^1$ with $N$ for $S=2.5$, $3$, $3.5$, and $4$. The 
numbers at the left of each graph are the $N \to \infty$ extrapolated values.}
\end{figure}

We now look at how the values of $\ka$ at the gapless points change with $N$
and see how the $N \rightarrow \infty$ values compare with those given by Eq.
(\ref{gaplesspts}). We take $\ka^1$ as an example. Figure 3 shows the behavior
for $S=2.5, 3, 3.5$, and 4. We find the $N \rightarrow \infty$ values by 
extrapolating the best fits obtained by fitting the data to even polynomials 
in $1/(2N)^2$ following Ref. \cite {nakamura}. All the data are for $N=2,3,4$
and 5 (for $S=2.5$ and 3).

The extrapolated values of $\ka^1$ in the $N \to \infty$ limit and the
corresponding values obtained from Eq. (\ref{gaplesspts}) (in parentheses) 
for $S=2.5$, $3.0$, $3.5$ and $4.0$ are given by $0.223 ~(0.200)$, 
$0.179 ~(0.167)$, $0.146 ~(0.143)$ and $0.124 ~(0.125)$ respectively.
Clearly, the agreement between the two sets of values gets better for 
larger values of the spin.

\section{\bf VI. Conclusions}

We have used three different techniques to find the gapless points of a 
dimerized spin-$S$ chain with a finite number of sites and
with a twisted boundary
condition. The first technique uses a tunneling approach which is expected to 
be valid in the limit $S \to \infty$, $\ka \to 0$ and $\ka S$ of order 1. 
Remarkably, we find that a quantum mechanical tunneling calculation reproduces
the same expressions for the locations of the gapless points and the gap as 
those obtained by more involved field theoretic techniques.

However, a direct numerical study of the gapless points shows a systematic
deviation from the tunneling results in the limit discussed above. It would be 
useful to know why the tunneling results differ systematically from the 
numerical results in this limit. One possible idea is to examine if an 
additional factor of $\pi$ appears in the fluctuation pre-factor of the 
tunneling amplitude as mentioned at the end of Sec. IV. 

In view of the discrepancy between the tunneling and numerical results, we have
presented a second analytical derivation of the gapless points which is based 
on a calculation of the matrix element between the two N\'eel states to lowest
order in powers of the Hamiltonian; this derivation is expected to become more
accurate as the number of sites becomes large. We find that the results 
obtained by this approach agree much better with the numerical results than 
the tunneling results, even in the limit $\ka \to 0$. It may be instructive to
understand in more detail why there is such a good agreement between this 
relatively simple analytical calculation and the numerical results.

One of the features of the numerical results shown above is that the N\'eel
state calculation always underestimates the values of $\ka$ which correspond to
gapless points, while all the time getting closer to the actual values with 
increasing $S$ (given $N$) or increasing $N$ (given $S$). This may indicate
positive corrections of order $1/N$ and $1/S$ to the formula obtained from the
N\'eel state calculation.

\begin{acknowledgments}
V.R.C. thanks CSIR, India for financial support. We thank DST, India for 
financial support under the project SP/S2/M-11/2000.
\end{acknowledgments}

\end{document}